\documentclass[aps,preprint,showpacs,nofootinbib]{revtex4}
\usepackage{epsfig} 
\usepackage{dcolumn}  

\graphicspath{{ps}}

\begin{document}


\newcommand{\gevc}{\ensuremath{\,{\mathrm{GeV}/c^2}}}
\newcommand{\mevc}{\ensuremath{\,{\mathrm{MeV}/c^2}}}
\newcommand{\gev} {\ensuremath{\,{\mathrm{GeV}}}}
\newcommand{\mev} {\ensuremath{\,{\mathrm{MeV}}}}
\newcommand{\ee}  {\ensuremath{e^+ e^- }}
\newcommand{\ds}  {\ensuremath{D^+_s}}
\newcommand{\dss} {\ensuremath{D^{*+}_s}}
\newcommand{\dso} {\ensuremath{D^-_{s1}}}
\newcommand{\dst} {\ensuremath{D^-_{s2}}}
\newcommand{\dm}  {\ensuremath{D^-}}
\newcommand{\dsp} {\ensuremath{D^{*+}}}
\newcommand{\dsm} {\ensuremath{D^{*-}}}
\newcommand{\RM}  {\ensuremath{ M_{\mathrm{recoil}} }}
\newcommand{\RMD} {\ensuremath{\Delta M_{\mathrm{recoil}} }}
\newcommand{\ks}  {\ensuremath{K_S^0}}
\newcommand{\dstbkm} {\ensuremath{\overline{D}{}^{*0}K^-}}
\newcommand{\dstmks} {\ensuremath{D^{*-}\ks}}

\newcommand{\yieldone}  {\ensuremath{184 \pm 19}}  
\newcommand{\yieldtwo}  {\ensuremath{105 \pm 14}}  
\newcommand{\yieldthree}  {\ensuremath{267 \pm 18}}  

\newcommand{\yieldfour}  {\ensuremath{45 \pm 8}}  
\newcommand{\resultsone}  {\ensuremath{ 4.01\pm 0.47(stat)}}  
\newcommand{\resultstwo}  {\ensuremath{ 3.84 \pm 0.83(stat)}}  

\newcommand{\results}  {\ensuremath{ 4.0 \pm 0.4(stat) \pm 0.4(sys)}}  

\preprint{\vbox{ \hbox{   }
                 \hbox{BELLE-CONF-0612}
}}

\title{ \quad\\[0.5cm] Measurement of the absolute branching fraction of
the $\mathbf{D_s^\pm}$ meson.}
\affiliation{Budker Institute of Nuclear Physics, Novosibirsk}
\affiliation{Chiba University, Chiba}
\affiliation{Chonnam National University, Kwangju}
\affiliation{University of Cincinnati, Cincinnati, Ohio 45221}
\affiliation{University of Frankfurt, Frankfurt}
\affiliation{The Graduate University for Advanced Studies, Hayama} 
\affiliation{Gyeongsang National University, Chinju}
\affiliation{University of Hawaii, Honolulu, Hawaii 96822}
\affiliation{High Energy Accelerator Research Organization (KEK), Tsukuba}
\affiliation{Hiroshima Institute of Technology, Hiroshima}
\affiliation{University of Illinois at Urbana-Champaign, Urbana, Illinois 61801}
\affiliation{Institute of High Energy Physics, Chinese Academy of Sciences, Beijing}
\affiliation{Institute of High Energy Physics, Vienna}
\affiliation{Institute of High Energy Physics, Protvino}
\affiliation{Institute for Theoretical and Experimental Physics, Moscow}
\affiliation{J. Stefan Institute, Ljubljana}
\affiliation{Kanagawa University, Yokohama}
\affiliation{Korea University, Seoul}
\affiliation{Kyoto University, Kyoto}
\affiliation{Kyungpook National University, Taegu}
\affiliation{Swiss Federal Institute of Technology of Lausanne, EPFL, Lausanne}
\affiliation{University of Ljubljana, Ljubljana}
\affiliation{University of Maribor, Maribor}
\affiliation{University of Melbourne, Victoria}
\affiliation{Nagoya University, Nagoya}
\affiliation{Nara Women's University, Nara}
\affiliation{National Central University, Chung-li}
\affiliation{National United University, Miao Li}
\affiliation{Department of Physics, National Taiwan University, Taipei}
\affiliation{H. Niewodniczanski Institute of Nuclear Physics, Krakow}
\affiliation{Nippon Dental University, Niigata}
\affiliation{Niigata University, Niigata}
\affiliation{University of Nova Gorica, Nova Gorica}
\affiliation{Osaka City University, Osaka}
\affiliation{Osaka University, Osaka}
\affiliation{Panjab University, Chandigarh}
\affiliation{Peking University, Beijing}
\affiliation{University of Pittsburgh, Pittsburgh, Pennsylvania 15260}
\affiliation{Princeton University, Princeton, New Jersey 08544}
\affiliation{RIKEN BNL Research Center, Upton, New York 11973}
\affiliation{Saga University, Saga}
\affiliation{University of Science and Technology of China, Hefei}
\affiliation{Seoul National University, Seoul}
\affiliation{Shinshu University, Nagano}
\affiliation{Sungkyunkwan University, Suwon}
\affiliation{University of Sydney, Sydney NSW}
\affiliation{Tata Institute of Fundamental Research, Bombay}
\affiliation{Toho University, Funabashi}
\affiliation{Tohoku Gakuin University, Tagajo}
\affiliation{Tohoku University, Sendai}
\affiliation{Department of Physics, University of Tokyo, Tokyo}
\affiliation{Tokyo Institute of Technology, Tokyo}
\affiliation{Tokyo Metropolitan University, Tokyo}
\affiliation{Tokyo University of Agriculture and Technology, Tokyo}
\affiliation{Toyama National College of Maritime Technology, Toyama}
\affiliation{University of Tsukuba, Tsukuba}
\affiliation{Virginia Polytechnic Institute and State University, Blacksburg, Virginia 24061}
\affiliation{Yonsei University, Seoul}
  \author{K.~Abe}\affiliation{High Energy Accelerator Research Organization (KEK), Tsukuba} 
  \author{K.~Abe}\affiliation{Tohoku Gakuin University, Tagajo} 
  \author{I.~Adachi}\affiliation{High Energy Accelerator Research Organization (KEK), Tsukuba} 
  \author{H.~Aihara}\affiliation{Department of Physics, University of Tokyo, Tokyo} 
  \author{D.~Anipko}\affiliation{Budker Institute of Nuclear Physics, Novosibirsk} 
  \author{K.~Aoki}\affiliation{Nagoya University, Nagoya} 
  \author{T.~Arakawa}\affiliation{Niigata University, Niigata} 
  \author{K.~Arinstein}\affiliation{Budker Institute of Nuclear Physics, Novosibirsk} 
  \author{Y.~Asano}\affiliation{University of Tsukuba, Tsukuba} 
  \author{T.~Aso}\affiliation{Toyama National College of Maritime Technology, Toyama} 
% ...
\collaboration{The Belle Collaboration}

\begin{abstract}
The $D_s^{\pm} \to K^{\pm}K^{\mp}\pi^{\pm}$ absolute branching fraction is measured
using $e^+e^-\to  D_s^{*\pm} D_{s1}^\mp(2536)$ events collected by the
Belle detector at the KEKB $e^+ e^-$ asymmetric energy collider. Using the ratio of yields when either the $D_{s1}$ or
$D_s^*$ is fully reconstructed, we find $\mathcal{B}(D_s^{\pm} \to K^{\pm}K^{\mp}\pi^{\pm})=(\results)$\%. 
\end{abstract}


\maketitle

\tighten

{\renewcommand{\thefootnote}{\fnsymbol{footnote}}}
\setcounter{footnote}{0}

Knowledge of \ds\ meson\footnote{Charge conjugation is implied through the
paper.}  absolute branching fractions is important for
normalization of many decays involving a \ds\ in a final state. The  poor
accuracy of the branching fraction ${\cal B}(\ds \to K^+ K^- \pi^+)=(5.2\pm 0.9)\%$
\cite{PDG} has been  a systematic limitation for some precise measurements. In
particular, the recent study of the $CP$ violation in $B^0 \to
D^{(*)\pm} \pi^{\mp}$ decays is restricted by the knowledge of the ratio
of two amplitudes that determine  the
$CP$-asymmetry~\cite{belle_dp,babar_dp}. The amplitude $B^0 \to D^{(*)+}
\pi^-$ can be calculated from the branching fraction of $B^0 \to
D_s^{(*)+} \pi^-$ decays assuming  factorization. On the other hand, the factorization hypothesis can be tested by measuring the ratio of  $B^0 \to D^{(*)-} \pi^+$ and $B^0 \to D^{(*)-}
D_s^+$ decays. Both  $\mathcal{B}(B^0 \to D_s^{(*)+} \pi^-)$
and $\mathcal{B}(B^0 \to D^{(*)-} D_s^+)$ measurements can be improved with better
accuracy in \ds\ absolute branching fractions.


Recently, the absolute branching fraction of $\ds \to \phi(\to K^+ K^-)\pi^+$ was
measured by the BaBar collaboration, which used partial and full
reconstruction of $B \to D^{(*)} D_s^{(*)+}$ decays~\cite{babar_dds}.
Another result obtained from a  $\sqrt{s}$-scan above $D^+_s D^-_s$
threshold was presented by the CLEO-c collaboration \cite{cleo_branching}.

In this paper we report on a measurement of the $\ds \to K^+ K^- \pi^+$ branching
fraction using two body \ee-continuum annihilation into a $\dss \dso(2536)$
final state. The analysis is based on $552.3\, \mathrm{fb}^{-1}$ of data
at the $\Upsilon(4S)$ resonance and nearby continuum, collected with
the Belle detector~\cite{belle} at the KEKB asymmetric energy storage
ring~\cite{kekb}.

\section{Method}

We use the partial reconstruction of the
process $\ee \to \dss D_{s1}^-$. In this analysis 4-momentum
conservation allows us to infer the 4-momentum of the undetected
part. The method used  was described in Ref.~\cite{DstDst} and applied to the
measurement of the $\ee \to D^{(*)+} D^{(*)-}$ cross sections.

Here we reconstruct the process $\ee \to \dss D_{s1}^-$ using two
different tagging procedures. The first one (denoted as the \dso\ tag)
includes the full reconstruction of the \dso\ meson via $\dso\to
\overline{D}{}^* K$ decay and observation of the photon from $\dss \to
\ds \gamma$, while the $D_s^+$ is not reconstructed. The measured
signal yield with the \dso\ tag is proportional to the branching fractions
of the reconstructed $\overline{D}{}^*$ modes. In the second procedure
(denoted as the \dss\ tag) we require full reconstruction of \dss\
through $\dss \to \ds \gamma$ and observation of the kaon from
$D_{s1}^{-}\to \overline{D}{}^{*} K$, but the $\overline{D}{}^{*}$ is
not reconstructed. Since the \ds\ meson is reconstructed in the
channel of interest, $\ds\to K^+ K^- \pi^+$, the signal yield measured
with the \dss\ tag is proportional to this \ds\ branching
fraction. The (efficiency-corrected) ratio of the two measured signal
yields is equal to the ratio of well-known $\overline{D}{}^{*}$
branching fractions and the branching fraction of the \ds.

In order to calculate reconstruction efficiencies and optimize event
selection criteria,  Monte Carlo (MC) signal events are generated and
simulated using a GEANT-based full simulator, including  initial
state radiation (ISR), and assuming no form-factors for \dss\ and \dso\ mesons.

To identify the signal we study the mass recoiling against the
reconstructed particle (or combination of particles) denoted as $X$. This recoil mass is defined  as:
\begin{eqnarray}
\RM(X) \equiv \sqrt{ {(E_{CM} - E_X)}^2 - P_X^2 },
\end{eqnarray}
where $E_X$ and $P_X$ are the center-of-mass (CM) energy and momentum of X,
respectively; $E_{CM}$ is the CM beam energy. We expect a peak in the \RM\ distribution at the
nominal mass of the recoil particle.

The resolution in \RM\ is   $\sim 50\mevc$ according to the MC, which is not sufficient to separate different final states, {\emph e.g.}  $\ds
D_{s1}^{-}$, $\dss D_{s1}^{-}$ and non-resonant $D_s^+ \overline{D} K$. To
disentangle the  contribution of these final states we use another
kinematic variable, the recoil mass difference \RMD:
\begin{eqnarray}
\RMD(D^{-}_{s1} \gamma) \equiv \RM(D^{-}_{s1})-\RM(D^{-}_{s1}\gamma), \\
\RMD(D^{*+}_s K) \equiv \RM(D^{*+}_s)-\RM(D^{*+}_s K).
\end{eqnarray}
In the {\dso\ tag} procedure the signal events make a narrow peak in
the $\RMD(D^{-}_{s1} \gamma)$ distribution at the nominal $\dss -\ds$ mass
difference with a resolution of $\sigma \sim 5\mevc$, dominated by the
$\gamma$ energy resolution according to the MC. The  $\RMD(D^{*+}_s K)$
spectrum peaks at the $D_{s1}^{-} - \overline{D}{}^{*}$ mass
difference with a resolution of $\sigma \sim 2\mevc$.

\section{Selection}

All primary charged tracks are required to be consistent with
originating from the interaction point. Charged kaon candidates are
identified using information from $dE/dx$ measurements in the central
drift chamber, aerogel Cherenkov counters and time-of-flight
system. No identification requirements are applied for pion
candidates. \ks\ candidates are reconstructed from $\pi^+
\pi^-$ pairs with an invariant mass within $15\mevc$ of the nominal
\ks\ mass. The distance between the two pion tracks at the \ks\ vertex
is required to be smaller than $1\,\mathrm{cm}$, the flight distance
in the plane perpendicular to the beam from the interaction point is
required to be greater than $0.1\,\mathrm{cm}$ and the angle between
the \ks\ momentum direction and decay path in this plane should be
smaller than $0.01\,\mathrm{rad}$. Photons are reconstructed in the
electromagnetic calorimeter as showers with energies above $50 \mev$
that are not associated with charged tracks. $\pi^0$ candidates are
reconstructed by combining pairs of photons with invariant masses
within $15\mevc$ of the nominal $\pi^0$ mass.

$D^0$ candidates are reconstructed using five decay modes: $K^-
\pi^+$, $K^- K^+$, $K^- \pi^- \pi^+ \pi^+$, $\ks \pi^+\pi^-$ and $K^-
\pi^+ \pi^0$. $D^+$ candidates are reconstructed using the $K^- \pi^+
\pi^+$ channel. \ds\ candidates are reconstructed using the $K^+ K^-
\pi^+$ decay mode. A $\pm 15\mevc$ mass window is used for all $D$
modes (approximately $2.5\sigma$ in each case) except for the $D^0
\to K^- \pi^- \pi^+ \pi^+$ where $\pm 10\mevc$ is applied and $D^0\to
K^- \pi^+ \pi^0$ with $\pm 20\mevc$ mass window.  All \ds, $D^+$ and
$D^0$ candidates are subjected to a mass and vertex constrained fit to
improve their momenta and thus the recoil mass resolution.

$D^*$ candidates are selected via $D^{*+} \to D^0 \pi^+$ and $D^{*0}
\to D^0 \pi^0$ decay modes with $\pm 2\mevc$ mass window. \dss\
candidates are reconstructed via $\ds \gamma$ channel with $\pm
10\mevc$ mass window. \dso\ is reconstructed in $\overline{D}{}^{*0}
K^-$ and $D^{*-} \ks$ decay modes.

In the case of multiple candidates, the candidate with the minimum value of
$\chi^2_{tot}$ ($\chi^2_{tot}=\chi^2_{M(D)} + \chi^2_{M(D^*)}$ for \dso\ tag
and $\chi^2_{tot}=\chi^2_{M(\ds)}$ for \dss\ tag, respectively) is chosen. Each
$\chi^2$ is defined as the square of the ratio of the deviation of the
measured mass from the nominal value and the corresponding resolution.

\section{Reconstruction}

\subsection{\dso\ tag}

As the ratio of $\dso\to\dstbkm$  and $\dso\to
\dstmks$ branching fractions is unknown, we perform the analysis for
these two channels separately and average the results.

\begin{figure}[htb]
\begin{center}
\epsfig{file=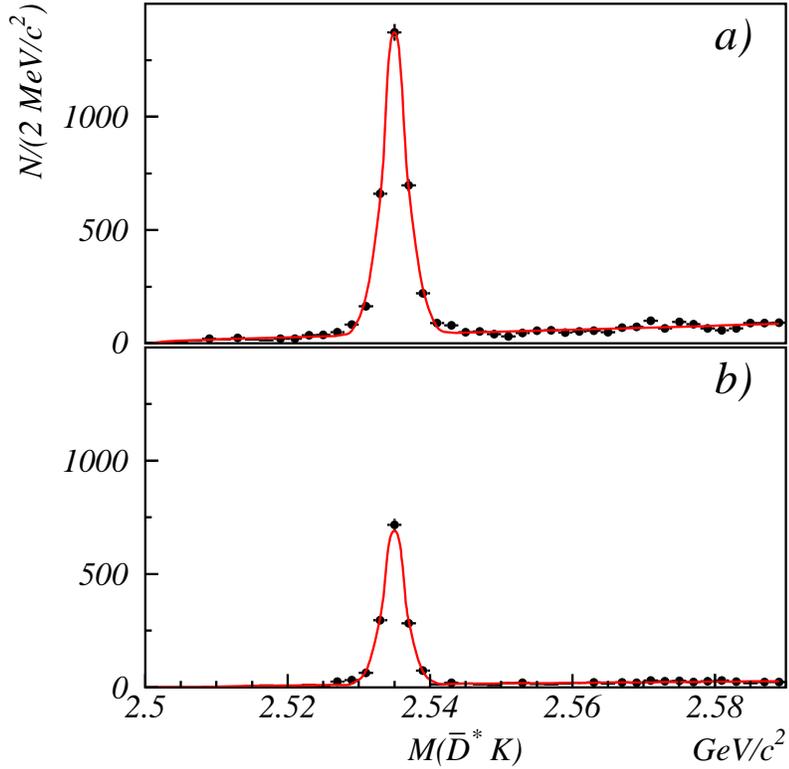,width=0.7\textwidth}
\end{center}
\caption{The distribution of invariant mass a) \dstbkm\ and b)
\dstmks. The solid curves represent the fit described in the text.}
\label {ds1.mds1}
\end{figure}

The \dstbkm\ and \dstmks\ mass spectra with a preselection requirement
on $\RM(\overline{D}{}^*K)~<3~\gevc$ are shown in
Fig.~\ref{ds1.mds1}. Clean peaks at the nominal \dso\ mass without
substantial combinatorial background are evident in both
distributions. We fit these distributions with a  sum of two Gaussians with  a common mean 
 and   the background parametrized with a  threshold
function:
\begin{equation}
Bg(m)=(A\cdot m + B )\sqrt{m-M^{PDG}_{\overline{D}{}^*}-M^{PDG}_{K}}.
\label{ds1.bg}
\end{equation}
The fit yields a Gaussian central value of $M=2535.1\pm
0.4\mevc$ for \dstbkm\ and $M=2535.0\pm 0.3\mevc$ for \dstmks\
channels, respectively, which is in  good agreement with both the PDG
and MC values.

\begin{figure}[htb]
\begin{center}
\epsfig{file=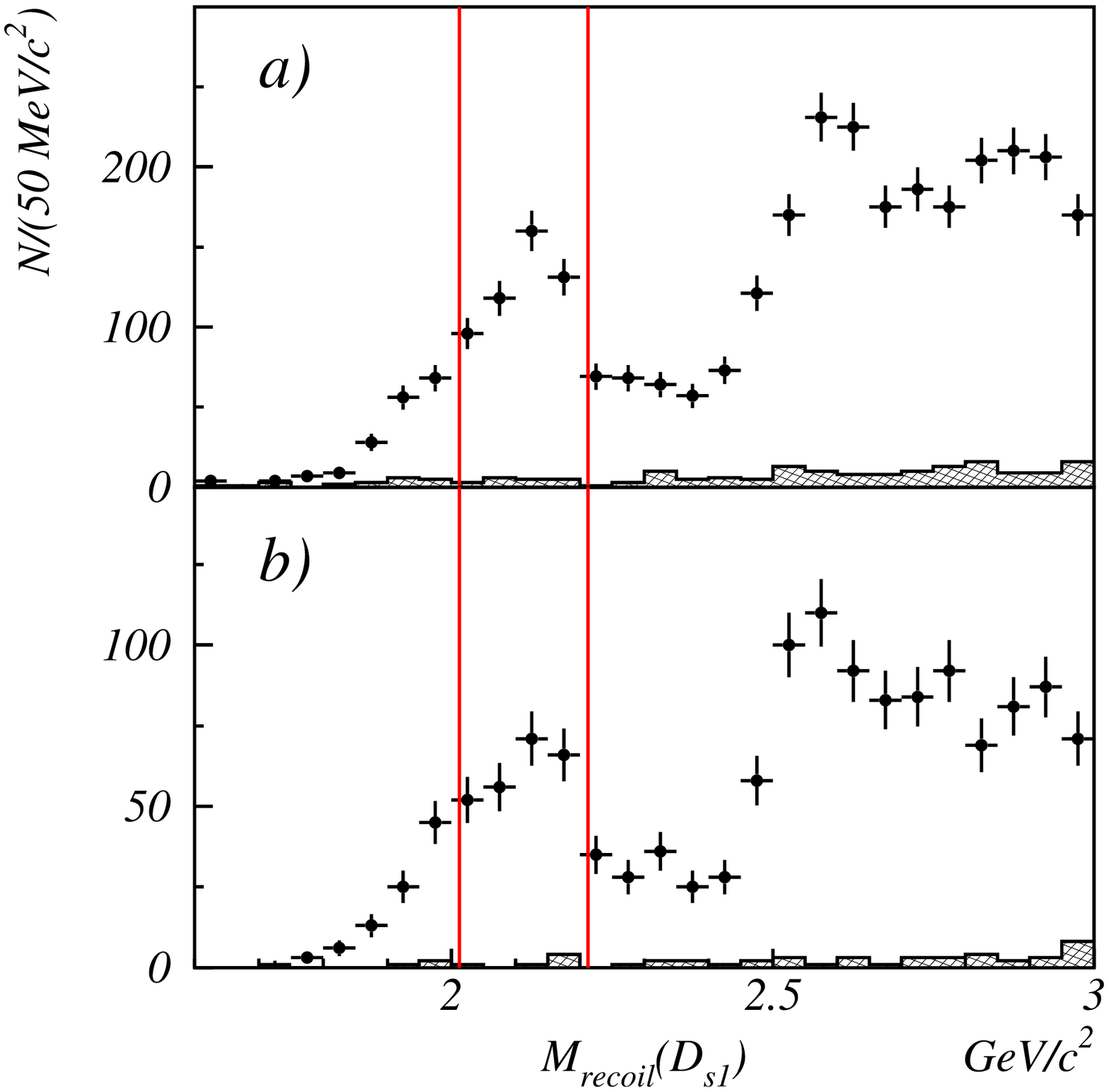 ,width=0.7\textwidth}
\end{center}
\caption{a) $\RM(\dso \to \dstbkm)$ distribution; b) $\RM(\dso \to
\dstmks )$ distribution.  Points with error bars correspond to the
events in the signal region. The \dso\ sideband contribution is shown as hatched
histogram. The  signal windows are indicated by the
vertical lines. }
\label{ds1.rm}
\end{figure}

The signal and sideband regions for \dso\ are defined as $|M(\dso)-2.5354|<0.005\gevc$ and $0.010\gevc<|M(\dso)-2.5354|<0.015\gevc$, respectively. The $\RM(\dso)$ spectra  are
shown in Fig.~\ref{ds1.rm}.  The significant excess in the region from
$1.9$ to $2.2 \gevc$ includes $\ee \to \ds \dso$ and $\ee \to \dss
\dso$ processes.   The small
excess around $2.3 \gevc$ could be interpreted as a 
contribution from the process $e^+e^-\to  D_{sJ}^+\dso$.  The excess
around $2.5 \gevc$ corresponds to the excited $D_s^{**+}=D_{s1}^+, ~
D_{s2}^+$ states.

We define the signal $\RM(\dso)$ window as
$|\RM(\dso)-2.1121|<0.100\gevc$ (Fig.~\ref{ds1.rm}). As the recoil mass resolution of $\sim50\mevc$ is
not sufficient to resolve  $\ee \to \ds \dso$ and $\ee \to \dss
\dso$ processes, to disentangle them we
reconstruct the $\gamma$ from $\dss \to  \ds\gamma$ decays. The reconstructed
\dso\ candidates from the signal $\RM(\dso)$ region are combined with
each photon in the event to calculate the recoil mass difference
$\RMD(\dso\gamma)$.

\begin{figure}[h!]
\begin{center}
\epsfig{file=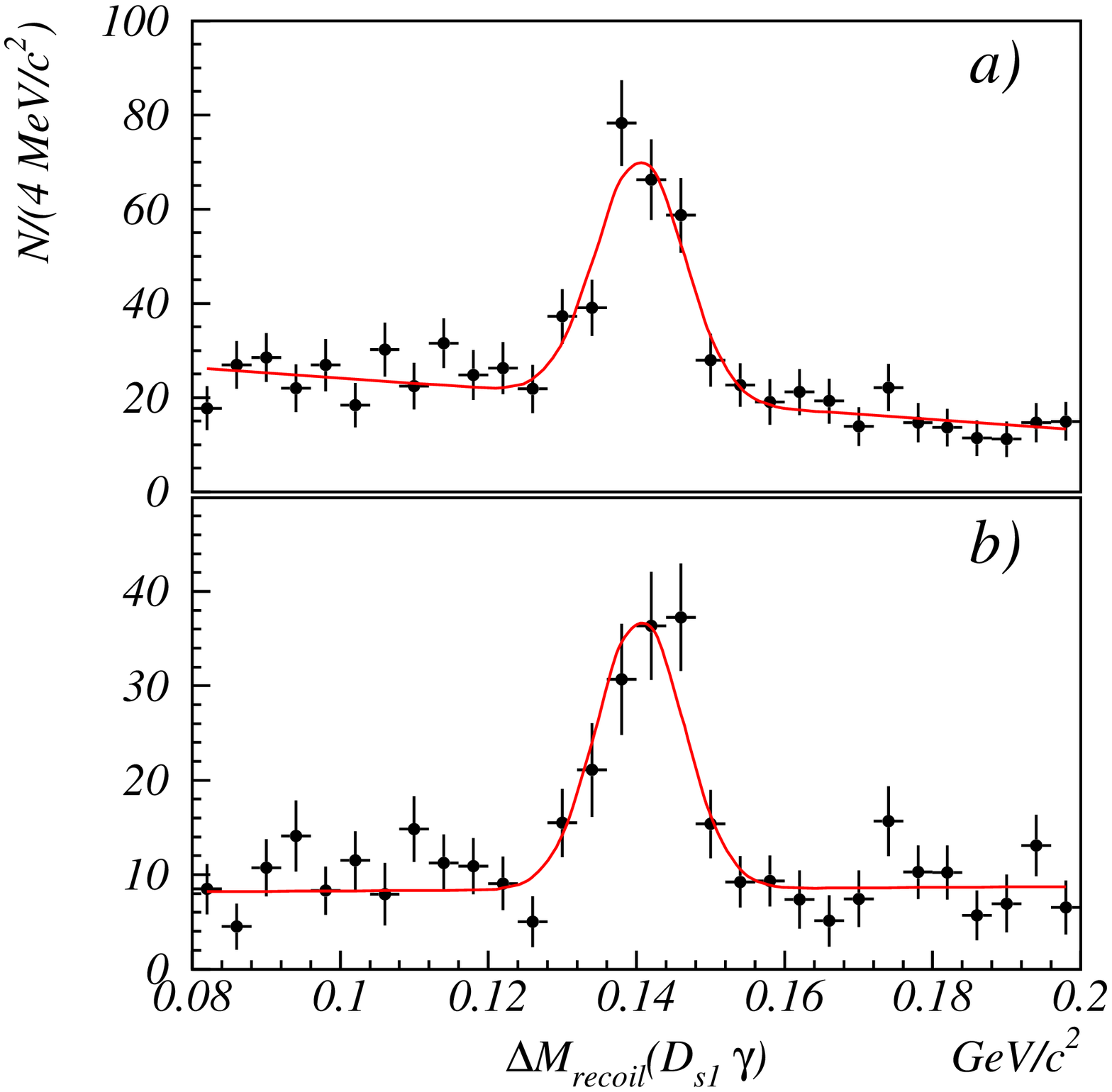,width=0.7\textwidth}
\end{center}
\caption{The signal yields in bins of $\RMD(\dso\gamma)$: a) for
the $\dso \to \dstbkm $ channel; b) for the $\dso \to \dstmks $
channel. The solid curves represent the fits described in the text.}
\label {ds1.binfit.results}
\end{figure}

To obtain the signal yield and disentangle the signal contribution
from the non-resonant $\dss \overline{D}{}^* K$ process, we perform fits to $\overline{D}{}^* K$
mass distributions in bins of $\RMD(\dso\gamma)$. The
$\RMD(\dso\gamma)$ bin width is chosen to be $4\mevc$. The signal
function is the sum of two  Gaussians with the mass and width fixed to the corresponding MC
values. The background is described with a threshold function
(Eq. \ref{ds1.bg}).
The signal yield for each bin  of the recoil mass difference for the $\dso\to\dstbkm$ and $\dso\to
\dstmks$ channels are plotted in Fig.~\ref{ds1.binfit.results}. 



The method used to extract the signal yield guarantees that  the
signal peak includes  only those events that contain both a \dso\
meson and a \dss\ decaying to $D_s\gamma$. Thus, the only possible
peaking background sources are  final states with extra pion(s)
({\emph e.g.} $\ee \to D^{*+}_s D_{s1}^- \pi^0$). The process with one
extra $\pi^0$ is suppressed due to isospin conservation. The possible
contribution of  $\ee \to \dso D_{sJ}^+(2460)$ followed by
$D_{sJ}^+(2460)\to \dss\pi^0$ decay, where the isospin violation
occurs in the $D_{sJ}$ decay, is suppressed by the recoil mass cut at the $4
\sigma$ level. The processes with two pions are also suppressed at
the $\sim3.5 \sigma$ level. Therefore, we conclude that the remaining
background is negligibly small; the extracted yields are 
dominated by the signal.

The distributions in Fig.~\ref{ds1.binfit.results} are fitted to the
sum of a signal Gaussian (with the mass and width fixed to the MC
values) and a linear background function. The fits yield
\yieldone\ signal events for $\dso \to\dstbkm$ and \yieldtwo\ events
for $\dso\to\dstmks$. To determine the reconstruction efficiency the
same procedure is applied to the MC signal sample.

\subsection{\dss\ tag}
\begin{figure}[htb]
\begin{center}
\epsfig{file=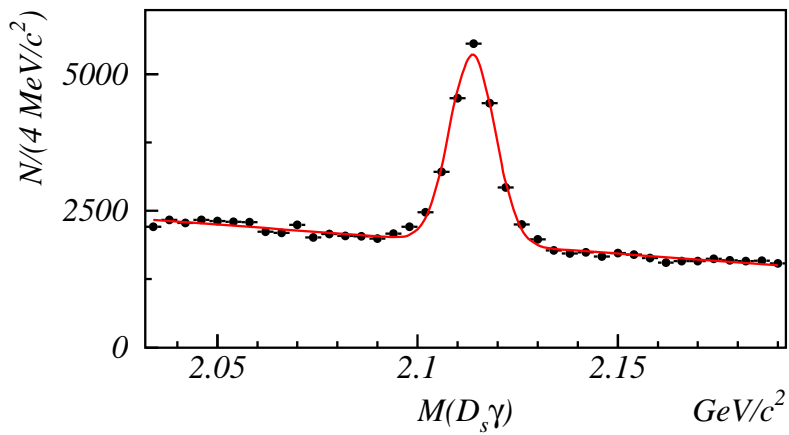,width=0.7\textwidth}
\end{center}
\caption{ The $\ds \gamma$ invariant mass distribution. The solid curve represents the fit described in the text.}
\label {dsst1.mdsst}
\end{figure}

In the \dss\ tag,  the \dss\ of  the  process $\ee\to\dss\dso$  is fully reconstructed.
The $\ds\gamma$ mass distribution is shown in
Fig.~\ref{dsst1.mdsst} and fitted with the sum of a signal Gaussian
and a linear background function. The obtained central value of the
Gaussian $M=2113.5\pm 0.4\mevc$ is somewhat larger than the  PDG value, but
 in  good agreement with the MC expectation, which shows a similar shift.  The signal region is
chosen to be $|M(\ds\gamma)-2.1121|<0.010\gevc$ while the sidebands
are defined as $0.015\gevc<|M(\ds\gamma)-2.1121|<0.025\gevc$.

\begin{figure}[htb]
\begin{center}
\epsfig{file=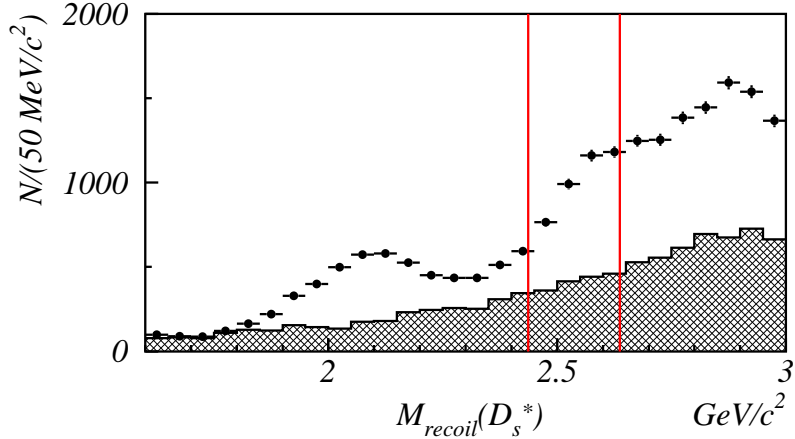,width=0.7\textwidth}
\end{center}
\caption{The $\RM(\dss)$ distribution. The events in the \dss\ signal region are shown as points
with error bars.  The \dss\ sidebands are superimposed as hatched
histogram. The  $\RM(\dss)$ signal window is indicated by
vertical lines. }
 \label {dsst1.rm}
\end{figure}

The $\RM(\dss)$ spectrum for the signal region is shown  in
Fig.~\ref{dsst1.rm}. A wide enhancement around $2.1\gevc$ is formed by
overlapping $D_s^-$ and $D_s^{*-}$ signals. The signal process $\ee\to
\dss \dso$ contributes to the bump around $2.55\gevc$ together with the
process $\ee \to  \dss\dst$. We define the signal region by the
requirement $|\RM(\dss)-2.5354|<0.100\gevc$ ($\sim2\sigma$).

\begin{figure}[h!]
\begin{center}
\epsfig{file=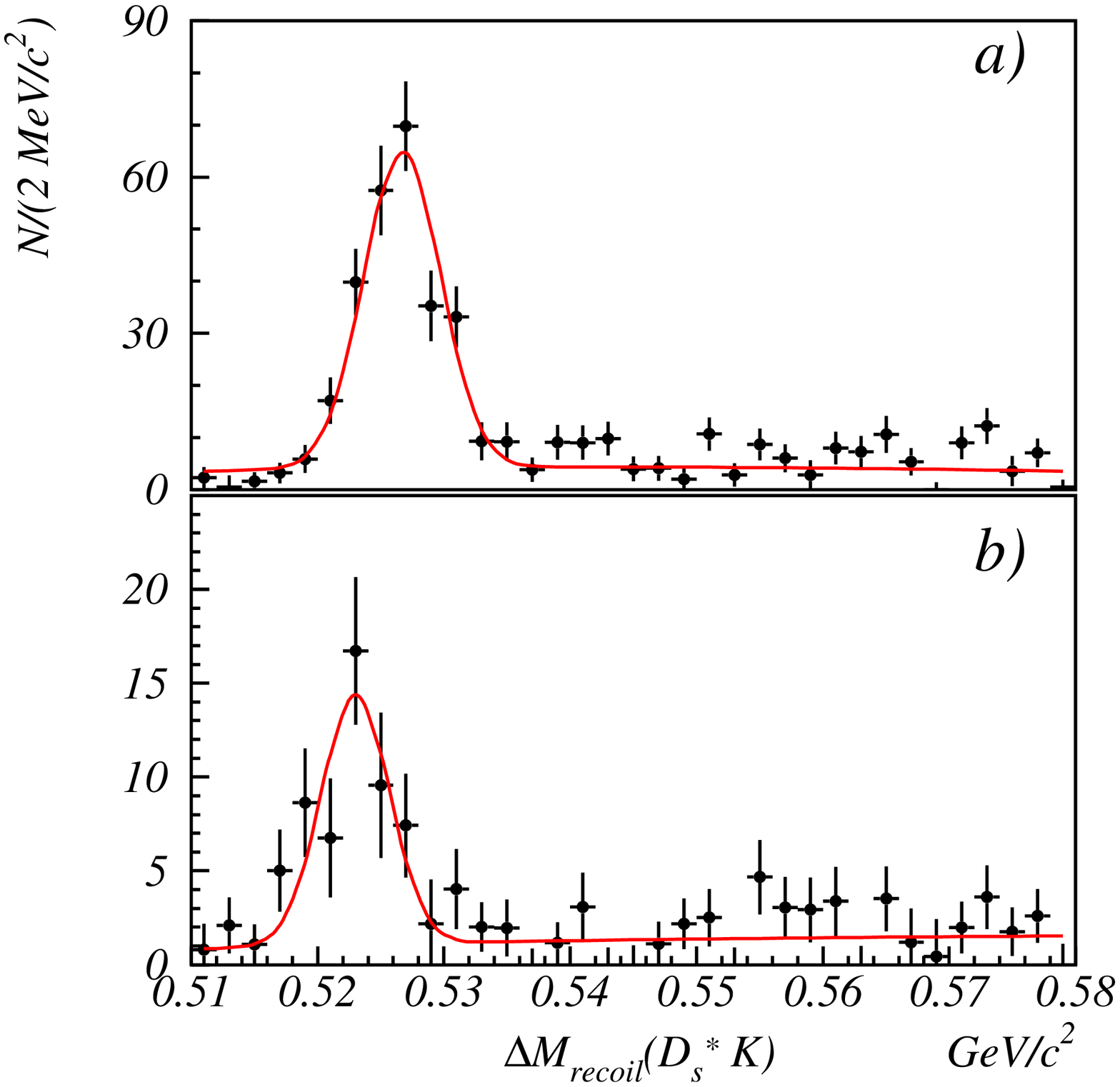,width=0.7\textwidth}
\end{center}
\caption{The $\ee\to\dss\dso$ yields for \dss\ tag in bins of $\RMD(\dss K)$: a) for
the $\dso \to \dstbkm $ channel; b) for the $\dso \to \dstmks$
channel. The solid curves show the fits described in the text.}
 \label {dsst1.binfit.results}
\end{figure}

To extract the $\ee \to  \dss\dso$ signal yield for \dss\ tag, we perform a fit to
the $\ds \gamma$ mass distributions in bins of  $\RMD(\dss K)$. The
signal function is a  Gaussian with its parameters fixed to the corresponding  MC
values. The background  is parametrized by a  linear function.
The yields returned by the fits for $\dso\to\dstbkm$ and $\dso\to
\dstmks$ channels are plotted in Fig.~\ref{dsst1.binfit.results}. We
fit the resulting distribution with the sum of a signal Gaussian (with
fixed mass and width) and a threshold function (Eq. \ref{ds1.bg}). The
fits yield \yieldthree\ signal events for the \dstbkm\ channel and
\yieldfour\ events for the \dstmks\ channel.

\section{Systematic errors}

The systematic uncertainties are summarized in Table \ref{systematics}.
\begin{table}[h!]
\caption{Contributions to the systematic error.}
\begin{tabular}{@{\hspace{0.3cm}}l@{\hspace{0.3cm}}||@{\hspace{0.3cm}}c@{\hspace{0.3cm}}}
\hline \hline
Source & Estimated error,  [\%] \\
\hline
Production angle     &  0.4  \\
$\overline{D}{}^*$ polarization   &  3.8  \\
ISR                  &  2.3  \\
Recoil mass cut      &  4.8  \\
Fit                  &  4.3  \\ 
Tracking/$\pi^0$ reconstr. efficiency             &  3.0  \\
${\cal B} (\overline{D}{}^{(*)})$ &  5.2  \\
MC statistics        &  2.3  \\
\hline
Total                &  10.1 \\
\hline
\hline
\end{tabular}

\label {systematics}
\end{table}

Different production and helicity angular distributions result in
different ratios of reconstruction efficiencies for the two tagging
procedures. A potential source of significant systematic error are the uncertainties, 
associated with reconstruction efficiencies of the kaon from the \dso\ decay
and the photon from the \dss, which strongly depend on the mother particle
polarizations. As the kaon and photon are, however,  reconstructed in both tags, the
systematic uncertainties related to the reconstruction of these
particles cancel out.  The remaining  systematic uncertainty in  the
efficiency ratio from  the production angle $\theta_{\mathrm{prod}}$ (polar angle
of the momentum of \dso\ meson) is estimated from MC simulation by
comparing the regions $|\cos(\theta_{\mathrm{prod}})|>0.5$ and
$|\cos(\theta_{\mathrm{prod}})|<0.5$ and found to be small 
($0.4\%$).

The  systematic error arising from unknown $\overline{D}{}^*$
polarizations contributes only to the uncertainty in the efficiency of
\dso\ tag and does not cancel out in the efficiency ratio. This
contribution is estimated from MC assuming  different $\overline{D}{}^*$
polarizations.


 The efficiency of the requirement on  recoil mass depends on the
fraction of events with an ISR photon as well as on the recoil mass
window cut. The error from the ISR photon is conservatively estimated
by changing the fraction of events containing ISR photons in the whole
MC sample by $30\%$. The resulting relative shift in ${\cal B}(\ds)$ is
found to be $2.3\%$. The error from the recoil mass cut is estimated
by varying the recoil mass window width in the range of
$80-120\mevc$. The difference in the calculated value of ${\cal B}(\ds)$ ($4.8\%$) is
conservatively considered as the systematic error.

The fit systematics is estimated assuming  different shapes for the
signal (single and double Gaussians) and for the background (different
order polynomials). The difference in the resulting branching fraction
is found to be $4.3$\% and is treated as the systematic uncertainty in  the
fit.

Other contributions come from the uncertainties in the track/$\pi^0$
reconstruction efficiencies, $\overline{D}{}^{(*)}$ branching fractions and MC statistics.

All of the systematic contributions   described above are added in quadrature to give $10.1$\%.

\section{Results}

Using the measured signal yields $N(\dss)$ and $N(\dso)$ with \dss\
and \dso\ tags, respectively, and taking into account the 
efficiency ratio ($\frac{ \epsilon(\dso)}{\epsilon(\dss)}=0.50$ for
\dstbkm\ and $\frac{ \epsilon(\dso)}{\epsilon(\dss)}=1.65$ for
\dstmks\ channels, respectively), we obtain the \ds\ absolute
branching fraction:
\begin{eqnarray}
  \mathcal{B} (\ds \to K^+ K^- \pi^+ )= \frac{N(\dss)}
  {N(\dso)}\cdot \frac{ \epsilon(\dso)}{\epsilon
  (\dss)}\cdot \mathcal{B}(\overline{D}{}^{(*)}),
\end{eqnarray}
where $\mathcal{B}(\overline{D}{}^{(*)})$ is the product of
$\overline{D}{}^*$ branching fraction and those of sub-decays. Separate calculations  for the \dstbkm\ and \dstmks\ channels
yield $\mathcal{B}(\ds \to K^+ K^- \pi^+)$ of (\resultsone)\% and
(\resultstwo)\%, respectively.

\section{Summary}

In summary, we report a measurement of the $\ds \to K^+ K^- \pi^+ $
branching fraction using a new method of  double tag partial
reconstruction. The branching fraction is measured separately in two channels
$\ee \to \dss\dso(\to \dstbkm)$ and $\ee \to \dss\dso(\to \dstmks)$. The average value is $\mathcal{B}(\ds \to K^+ K^-
\pi^+)=(\results$)\%.

\end{document}